\documentclass[aps,pra,twocolumn,superscriptaddress]{revtex4}
\usepackage{graphicx}
\usepackage{subfigure}
\usepackage{epstopdf}
\usepackage{amsmath}
\usepackage{amssymb}
\usepackage{amsfonts}
\usepackage{mathrsfs}
\usepackage{theorem}
\usepackage{bm}
\usepackage{url}
\usepackage[T1]{fontenc}
\usepackage{csquotes}
\MakeOuterQuote{"}

\usepackage{algorithm}
\usepackage{algorithmicx}
\usepackage{algpseudocode}

\usepackage{dcolumn}
\usepackage{color}

\definecolor{ngreen}{rgb}{0.2,0.6,0.2}

\definecolor{ngold}{rgb}{0.7,0.6,0.2}

\def\vec#1{\mathbf{#1}} 

\newcommand{\tr}{\operatorname{tr}}

\newcommand{\rmi}{\mathrm{i}}
\newcommand{\rme}{\mathrm{e}}

\newcommand{\one}{\overline{1}}
\newcommand{\zero}{\overline{0}}

\newcommand{\be}{\begin{equation}}
\newcommand{\ee}{\end{equation}}
\newcommand{\ba}{\begin{align}}
\newcommand{\ea}{\end{align}}

\def\<{\langle}  
\def\>{\rangle}  
\newcommand{\ket}[1]{| #1\>}
\newcommand{\bra}[1]{\< #1|}

\def\outer#1#2{|#1\>\<#2|}       

\newcommand{\cF}{\mathcal{F}}

\def\eqref#1{\textup{(\ref{#1})}}  
\newcommand{\eref}[1]{Eq.~\textup{(\ref{#1})}}

\newcommand{\esref}[1]{Eqs.~\textup{(\ref{#1})}}

\newcommand{\fref}[1]{Fig.~\ref{#1}}
\newcommand{\Fref}[1]{Figure~\ref{#1}}
\newcommand{\fsref}[1]{Figs.~\ref{#1}}

\newcommand{\tref}[1]{Table~\ref{#1}}

\newcommand{\sref}[1]{Sec.~\ref{#1}}

\newcommand{\cref}[1]{Conjecture~\ref{#1}}
\newcommand{\Cref}[1]{Conjecture~\ref{#1}}

\newcommand{\rcite}[1]{Ref.~\cite{#1}}
\newcommand{\rscite}[1]{Refs.~\cite{#1}}

\begin{document}

\title{Deterministic realization of  collective measurements via photonic quantum walks}
\author{Zhibo Hou}
\affiliation{Key Laboratory of Quantum Information,University of Science and Technology of China, CAS, Hefei 230026, P. R. China}
\affiliation{Synergetic Innovation Center of Quantum Information and Quantum Physics, University of Science and Technology of China, Hefei 230026, P. R. China}
\author{Jun-Feng Tang}
\affiliation{Key Laboratory of Quantum Information,University of Science and Technology of China, CAS, Hefei 230026, P. R. China}
\affiliation{Synergetic Innovation Center of Quantum Information and Quantum Physics, University of Science and Technology of China, Hefei 230026, P. R. China}
\author{Jiangwei Shang}
\affiliation{Naturwissenschaftlich-Technische Fakult{\"a}t, Universit{\"a}t Siegen, Siegen  57068, Germany}
\affiliation{Beijing Key Laboratory of Nanophotonics and Ultrafine Optoelectronic Systems, School of Physics, Beijing Institute of Technology, Beijing 100081, China}
\author{Huangjun Zhu}
\email{zhuhuangjun@fudan.edu.cn}
\affiliation{Institute for Theoretical Physics, University of Cologne,  Cologne 50937, Germany}

\affiliation{Department of Physics and Center for Field Theory and Particle Physics, Fudan University, Shanghai 200433, China}

\affiliation{Institute for Nanoelectronic Devices and Quantum Computing, Fudan University, Shanghai 200433, China}

\affiliation{State Key Laboratory of Surface Physics, Fudan University, Shanghai 200433, China}

\affiliation{Collaborative Innovation Center of Advanced Microstructures, Nanjing 210093, China}

\author{Jian Li}
\affiliation{Institute of Signal Processing Transmission, Nanjing University of Posts and Telecommunications, Nanjing 210003, China}

\affiliation{Key Lab of Broadband Wireless Communication and Sensor Network Technology, Nanjing University of Posts and Telecommunications, Ministry of Education, Nanjing 210003, China}

\author{Yuan~Yuan}
\affiliation{Key Laboratory of Quantum Information,University of Science and Technology of China, CAS, Hefei 230026, P. R. China}
\affiliation{Synergetic Innovation Center of Quantum Information and Quantum Physics, University of Science and Technology of China, Hefei 230026, P. R. China}
\author{Kang-Da Wu}
\affiliation{Key Laboratory of Quantum Information,University of Science and Technology of China, CAS, Hefei 230026, P. R. China}
\affiliation{Synergetic Innovation Center of Quantum Information and Quantum Physics, University of Science and Technology of China, Hefei 230026, P. R. China}

\author{Guo-Yong Xiang}
\email{gyxiang@ustc.edu.cn}
\affiliation{Key Laboratory of Quantum Information,University of Science and Technology of China, CAS, Hefei 230026, P. R. China}
\affiliation{Synergetic Innovation Center of Quantum Information and Quantum Physics, University of Science and Technology of China, Hefei 230026, P. R. China}
\author{Chuan-Feng Li}
\affiliation{Key Laboratory of Quantum Information,University of Science and Technology of China, CAS, Hefei 230026, P. R. China}
\affiliation{Synergetic Innovation Center of Quantum Information and Quantum Physics, University of Science and Technology of China, Hefei 230026, P. R. China}
\author{Guang-Can Guo}
\affiliation{Key Laboratory of Quantum Information,University of Science and Technology of China, CAS, Hefei 230026, P. R. China}
\affiliation{Synergetic Innovation Center of Quantum Information and Quantum Physics, University of Science and Technology of China, Hefei 230026, P. R. China}

\begin{abstract}
Collective measurements on identically prepared quantum systems can extract more information than local measurements, thereby enhancing information-processing efficiency. Although this nonclassical phenomenon has been known for two decades, it has remained a challenging task to demonstrate the advantage of collective measurements in experiments.
Here we introduce a general recipe for performing  deterministic collective measurements on two identically prepared qubits based on quantum walks. Using photonic quantum walks, we realize experimentally  an optimized  collective measurement  with  fidelity 0.9946 without post selection.  As an application, we achieve the highest tomographic efficiency  in qubit state tomography to date.  Our work offers an effective recipe for beating the precision limit of local measurements in quantum state tomography and  metrology. In addition, our study  opens an avenue for harvesting the power of collective measurements in quantum information processing and for exploring the  intriguing physics behind this power.

\end{abstract}

\date{\today}
\maketitle

Quantum measurements are the key for  extracting information from quantum systems and for connecting the quantum world with the classical world. Understanding the power and limitation of  measurements is of paramount importance not only to foundational studies, but also to many  applications, such as quantum tomography,  metrology, and  communication \cite{Hels76book, Hole82book, BrauC94,PariR04Book, LvovR09,RiedBLH10, GiovLM11,SzczBD16}. An intriguing phenomenon predicted by quantum theory is that collective measurements on identically prepared quantum systems may extract more information
than local measurements on individual  systems, thereby leading to higher tomographic efficiency and precision 
\cite{MassP95,GisiP99, VidaLPT99,BagaBGM06S,Zhu12the,ZhuH17U}. Recently the significance of collective measurements for multiparameter quantum metrology was also theorectically recognized \cite{VidrDGJ14,RoccGMS17} and experimentally demonstrated with probalistic Bell measurements as a proof of principle \cite{RoccGMS17}. 
This nonclassical phenomenon is owing  to entanglement in the quantum measurements instead of  quantum states.  It is closely tied to the phenomenon of "nonlocality without entanglement"~\cite{BennDFM99}.
In addition, collective measurements are very useful in numerous other tasks, such as  distilling entanglement \cite{BennBPS96P}, enhancing nonlocal correlations \cite{LianD06}, and  detecting quantum change point  \cite{SentBCC16}.
However, demonstrating the advantage of collective measurements  in experiments has remained a daunting task. This is  because most optimized protocols  entail generalized entangling measurements on many identically prepared quantum systems, which are very difficult to realize deterministically.

Here  we introduce a general method for performing deterministic collective measurements on two identically prepared qubits based on quantum walks, 
which extends the method for performing generalized measurements on a single qubit only  \cite{KurzW13,BianLQZ15,ZhaoYKX15}. By devising 
 photonic quantum walks, we realize experimentally  a highly efficient collective measurement highlighted  in  \rscite{VidaLPT99,Zhu12the,ZhuH17U}. 
As an application,  we realize, for the first time, qubit state tomography with deterministic collective measurements. The  protocol we implemented  is significantly more efficient than  local measurements  commonly employed in most experiments.
Moreover, it can achieve  near-optimal performance over all two-copy collective measurements
with respect to various figures of merit without using  adaptive measurements. 
Such high efficiency  demonstrates the main advantage of   collective measurements over separable measurements.
Here, we encode the two qubits in the two degrees of freedom of a single photon \cite{EnglKW01,Fior04deterministic,Barr08beating,Rubi17experimental}, but our method for performing collective measurements can be generalized to two-photon two-qubit states by combining the technique of quantum joining \cite{vite13joining} or teleportation~\cite{Ahne06all}.

\begin{figure*}
 	\center{\includegraphics[scale=0.85]{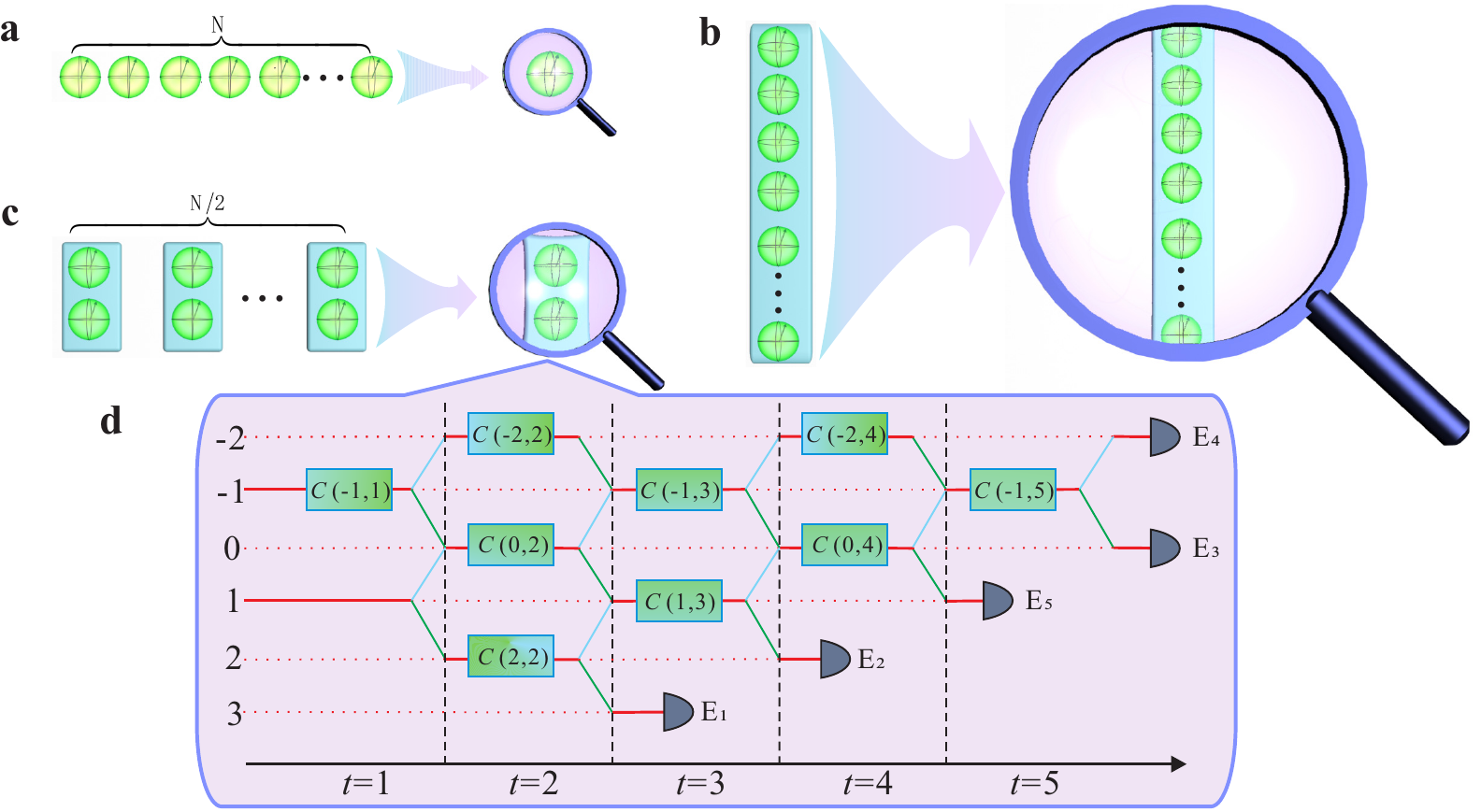}}
 	\caption{\label{fig:coll_theory} Individual and collective measurements. \textbf{a}: Repeated individual measurements. \textbf{b}: Single $N$-copy collective measurement. \textbf{c}: Repeated two-copy collective measurements. 	\textbf{d}:
 	Realization of the collective SIC-POVM defined in \esref{eq:CollectiveSIC} and \eqref{eq:QubitSIC} using five-step  quantum walks. The  coin qubit and the walker in positions 1 and $-1$ are taken as the two-qubit system of interest, while the other positions of the walker act as an ancilla.  Site-dependent  coin operators $C(x,t)$  are specified in the Methods section. Five detectors  $E_1$ to $E_5$  correspond to the five outcomes of the collective SIC-POVM. 									
 }
 \end{figure*}

\bigskip
\noindent\textbf{Results}\\
\noindent\textbf{Optimized collective measurements.} In quantum theory, a measurement is usually represented by a positive-operator-valued measure (POVM), which is composed of a set of positive operators that sum up to the identity.
In traditional quantum information processing,  measurements are performed on individual quantum systems one by one, which often cannot extract information efficiently. Fortunately, quantum theory allows us to perform  collective measurements on identically prepared  quantum systems in a way that has no classical analog, as illustrated in
 \fref{fig:coll_theory}.
 


In the case of a qubit, a special  two-copy collective POVM was highlighted in \rscite{VidaLPT99,Zhu12the,ZhuH17U}, which consists of five POVM elements,
\begin{equation}\label{eq:CollectiveSIC}
E_j =\frac{3}{4}(|\psi_j\>\<\psi_j|)^{\otimes 2}, \quad E_5=|\Psi_-\>\<\Psi_-|,
\end{equation}
where $|\Psi_-\>=\frac{1}{\sqrt{2}}(|01\>-|10\>)$ is the singlet, which is maximally entangled, and $|\psi_j\>$ for $j=1,2,3,4$ are qubit states that form a
symmetric informationally complete POVM (SIC-POVM), that is, $|\<\psi_j|\psi_k\>|^2=(2\delta_{jk}+1)/3$ \cite{Zaun11,ReneBSC04}. Geometrically, the Bloch vectors of the four states $|\psi_j\>$ form a regular tetrahedron inside the Bloch sphere. For concreteness, here we choose
\begin{equation}\label{eq:QubitSIC}
\begin{aligned}
\ket{\psi_1}=&\ket{0},\qquad \ket{\psi_2}=\frac{1}{\sqrt{3}}(\ket{0}+\sqrt{2}\ket{1}), \\ \ket{\psi_3}=&\frac{1}{\sqrt{3}}(\ket{0}+\rme^{\frac{2\pi}{3}\rmi}\sqrt{2}\ket{1}),\\ \ket{\psi_4}=&\frac{1}{\sqrt{3}}(\ket{0}+\rme^{-\frac{2\pi}{3}\rmi}\sqrt{2}\ket{1}).
\end{aligned}
\end{equation}
The POVM defined by \esref{eq:CollectiveSIC} and \eqref{eq:QubitSIC} is referred to as the collective SIC-POVM henceforth. If this POVM is performed on the two-copy state $\rho^{\otimes2}$, then the probability of obtaining outcome $j$ is given by $p_j=\tr(\rho^{\otimes 2}E_j)$. 



The collective SIC-POVM is distinguished because it is optimal 
in extracting information from a pair of identical qubits \cite{MassP95,VidaLPT99}. It is universally Fisher symmetric in the sense of providing uniform and maximal Fisher information on all parameters that characterize the quantum states of interest~\cite{Zhu12the, LiFGK16,ZhuH17U}. Moreover, it is the unique such POVM with no more than five outcomes.
Consequently, the collective SIC-POVM is significantly more efficient than any local measurement in many quantum information processing tasks, including  tomography and metrology. Moreover, its high tomographic efficiency is achieved without using adaptive measurements, which is impossible for local measurements.

\begin{figure*}[t]	
	\center{\includegraphics[scale=0.45]{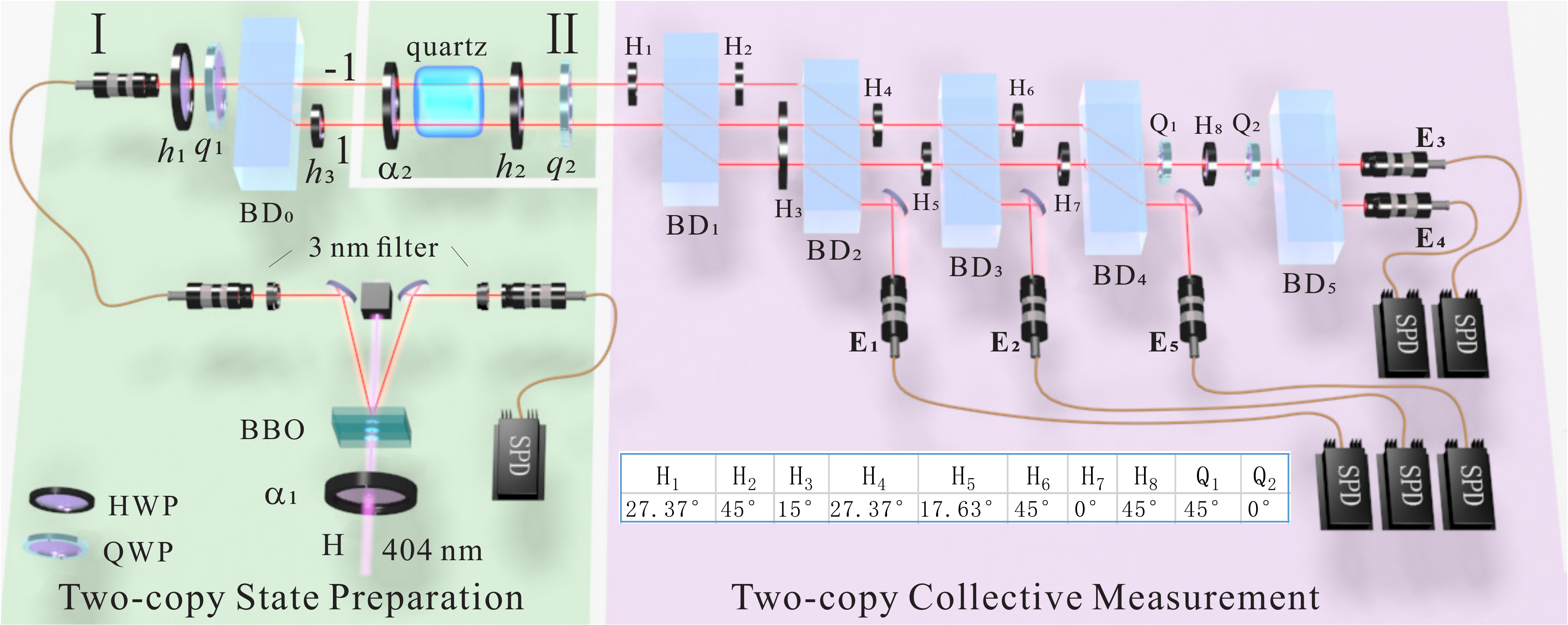}}
	\caption{\label{fig: exp setup}  Experimental setup
	for realizing the collective SIC-POVM. The setup consists of two modules designed for  two-copy state preparation (green) and two-copy collective measurement (purple), respectively. In the two-copy state preparation module, submodule~\uppercase\expandafter{\romannumeral1} prepares the first copy  (walker qubit) in the path degree of freedom;
		submodule~\uppercase\expandafter{\romannumeral2} prepares	
		the second copy (coin qubit) in the polarization degree of freedom.
		The two-copy collective-measurement module performs the collective SIC-POVM  via photonic quantum walks as illustrated in \fref{fig:coll_theory}d.  Here beam displacers  (BDs) are used to realize the conditional translation operator $T$.  Combinations of half wave plates (HWPs) and quarter wave plates (QWPs) with rotation angles specified in the embedded table are used to realize site-dependent  coin operators $C(x,t)$.
		Five single-photon detectors (SPDs) $E_1$ to $E_5$  correspond to the five outcomes of the collective SIC-POVM.				
	}
\end{figure*}

\bigskip
\noindent\textbf{Realization of the collective SIC-POVM via quantum walks.}
Recently, discrete quantum walks were proposed as a recipe for implementing general POVMs on a single qubit \cite{KurzW13}, which have been demonstrated in experiments \cite{BianLQZ15,ZhaoYKX15}. In a one-dimensional discrete quantum walk, the system state is characterized by two degrees of freedom $|x,c\>$, where $x=\cdots, -1,0,1,\cdots$ denotes the walker position, and $c=0,1$ represents the coin state. The dynamics of each step is described by a unitary transformation of the form $U(t)=TC(t)$, where
\begin{align}
  T = \sum_x \ket{x+1,0}\bra{x,0} + \ket{x-1,1}\bra{x,1}
  \end{align}
is the conditional translation operator,
and $C(t)=\sum_x |x\>\<x|\otimes C(x,t)$  with $C(x,t)$ being site-dependent coin operators. A general POVM on a qubit can be realized by engineering the coin operators $C(x,t)$ followed by measuring the walker position after certain steps. However, little is known in the literature on realizing POVMs on higher dimensional systems. Here we propose a general method for extending the capabilities of quantum walks. For concreteness, we illustrate our approach with the collective SIC-POVM.

 To realize the collective SIC-POVM using quantum walks, the coin qubit and the walker in positions 1 and $-1$ are taken as the two-qubit system of interest, while the other positions of the walker act as an ancilla. With this choice,
 the collective SIC-POVM can be realized with five-step  quantum walks, as illustrated in \fref{fig:coll_theory}d and discussed in more details in the supplement. Here the nontrivial coin operators $C(x,t)$  are specified in the Methods section. 
 The five detectors $E_1$ to $E_5$ marked in the figure correspond to the five POVM elements specified in \esref{eq:CollectiveSIC} and \eqref{eq:QubitSIC}. Moreover, this proposal can be implemented using photonic quantum walks, as illustrated in \fref{fig: exp setup} (see also Supplementary \fref{fig:coll_theorySupp}).


\bigskip
\noindent\textbf{Experimental setup.}
The experimental setup for realizing the collective SIC-POVM and its application in quantum state tomography
is presented in \fref{fig: exp setup}. The setup is composed of  two modules designed for  two-copy state preparation and collective measurements, respectively.

The two-copy collective measurement module performs the collective SIC-POVM based on quantum walks, as illustrated in \fref{fig:coll_theory}d (cf.~Supplementary \fref{fig:coll_theorySupp}). Here the conditional translation operator $T$ is realized by  interferometrically stable beam displacers  (BDs) \cite{Obr03demonstration,Rah13direct,Rab17entanglement,Sha15strong}, which displace the component with horizontal polarization (H)  away from the component  with vertical polarization (V). The coin operators $C(x,t)$ are realized by  suitable combinations of half wave plates (HWPs) and quarter wave plates (QWPs), with rotation angles specified in  the table embedded in \fref{fig: exp setup}.

In the two-copy state-preparation module,  we first prepare copy 1 in the path degree of freedom, i.e., the walker qubit encoded in positions 1 and $-1$ (see the green region \uppercase\expandafter{\romannumeral1} in \fref{fig: exp setup}). A pair of 1-mm-long BBO crystals with optical axes perpendicular to each other, cut for type-\uppercase\expandafter{\romannumeral1} phase-matched spontaneous parametric down-conversion (SPDC) process, is pumped by a 40-mW H-polarized beam at 404~nm. The  polarization state of the beam is prepared as  $\cos 2\alpha_1\ket{H}+\sin 2\alpha_1\ket{V}$ when the deviation angle of the HWP at 404 nm is set at $\alpha_1$. After the SPDC process,
 a pair of photons with wave length $\lambda=808$~nm is created in the state of $\sin 2\alpha_1\ket{HH}+\cos 2\alpha_1\ket{VV}$ \cite{Kwia99ultrabright}. The two photons pass through two interference filters whose FWHM (full width at half maximum) is 3 nm, resulting in a coherence length of 270$\lambda$. One photon is detected by a single-photon detector acting as a trigger. After tracing out this photon, the other photon is prepared in the state $\sin^2 2\alpha_1\outer{H}{H}+\cos^2 2\alpha_1\outer{V}{V}$, whose purity is controlled by $\alpha_1$. Two HWPs (not shown in \fref{fig: exp setup}) at the input and output ports of the single-mode fiber are used to maintain the polarization state of the photon.
 After passing
 a HWP  and a QWP  with deviation angles $h_1, q_1$, the photon is prepared in the desired
 state $\rho$.  To encode the polarization state into the path degree of freedom,  BD$_0$ is used to displace the H-component into path 1, which is  4-mm away from the V-component in path $-1$; then  a HWP with deviation angle  $h_3=45^\circ$ is placed in  path 1.  The resulting photon is described by the state $\rho\otimes\outer{V}{V}$.

 Then we encode the second copy of $\rho$ into the polarization degree of freedom  (coin qubit) using two HWPs, a quartz crystal with a decoherence length of 385$\lambda$, and a QWP (see the green region \uppercase\expandafter{\romannumeral2} in \fref{fig: exp setup}). The first HWP with  rotation angle  $\alpha_2$ and the quartz crystal prepare the polarization state $\sin^2 2\alpha_2\outer{H}{H}+\cos^2 2\alpha_2\outer{V}{V}$ with desired purity. Then the direction of the Bloch vector of the polarization state is adjusted by a HWP and a QWP with deviation angles  $h_2$ and $q_2$. In this way,  we can prepare the desired two-copy state $\rho\otimes \rho$, the first copy of which is encoded in the path degree of freedom, while the second one  in the polarization degree of freedom.

 The two-copy state-preparation module described above  is capable of preparing any two-copy state. Next,  the two-copy state $\rho\otimes \rho$ is sent into the two-copy collective-measurement module, which performs the collective SIC-POVM based on quantum walks, as described before. It is worth pointing out that the collective SIC-POVM can also be applied to measure arbitrary two-qubit states, although  we focus on two-copy qubit states in this work.

\begin{figure}
	\centering
\includegraphics[width=0.4\textwidth]{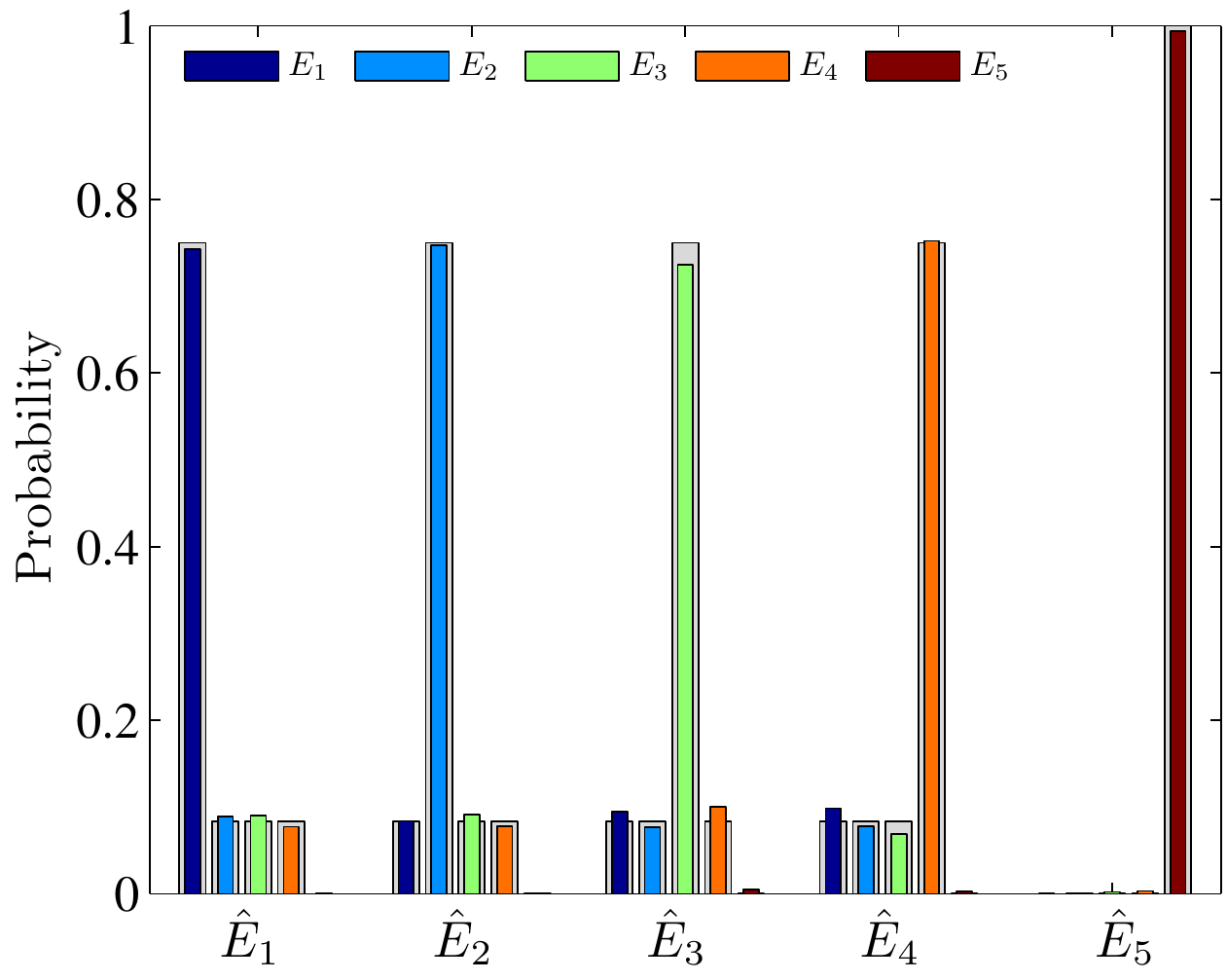}
\caption{\label{fig: probability distribution}Experimental verification of the  collective SIC-POVM realized. Here each $\hat{E}_i$ for $i=1,2,3,4,5$ denotes an  input state which corresponds to the  POVM element $E_i$ of the collective SIC-POVM after normalization. Each input state is prepared and measured 100000 times.
The frequencies of obtaining the five outcomes  are plotted using different colors; here the error bars are too small to be visible. For comparison, the probabilities in the ideal scenario are plotted in grey shadow.  }
\end{figure}

\bigskip
\noindent \textbf{Experimental verification of the collective SIC-POVM.}
To verify the experimental implementation of the collective SIC-POVM, we  took the conventional method of measuring the probability distributions after preparing the input states as normalized POVM elements, i.e., $\hat{E}_i=E_i/\tr(E_i)$ for $i=1,\cdots,5$.
These input states  can be prepared by  choosing proper rotation angles $\alpha_1,h_1,q_1,h_3,\alpha_2,h_2,q_2$ as specified in the supplement. The measurement probability distributions are shown in \fref{fig: probability distribution}, which  agree very  well with the theoretical prediction.

To accurately characterize the POVM elements that were actually realized, we then performed quantum measurement tomography. Overall, 36 input states,  the tensor products of the six eigenstates of three Pauli operators, were prepared and sent to the collective-measurement module, with each setting repeated 35000 times.
Then
 the five POVM elements were estimated from the  measurement statistics using the maximum likelihood  method developed in \rcite{Fiur01maximum}. The fidelities of the five  POVM elements estimated  are $0.9991\pm0.0001$,   $0.9979\pm0.0007$,        $0.9870\pm0.0008$,     $0.9927\pm0.0002$ and  $0.9961\pm0.0002$, respectively;  the overall fidelity of the POVM  (cf.~the Methods section) is
  $0.9946\pm0.0002$. Here the error bars denote the standard deviations of  100 simulations from Poisson statistics. Such high fidelities demonstrate that the collective SIC-POVM was realized with very high quality. Detailed information about the five reconstructed POVM elements can be found in the
  supplement.

\bigskip
\noindent \textbf{quantum state tomography with the  collective SIC-POVM.}
The experimental realization of the collective SIC-POVM enables us to achieve unprecedented efficiency in quantum state tomography. In this section we demonstrate the tomographic significance of the collective SIC-POVM and the power of collective measurements.

\begin{figure*}
	\center{\includegraphics[scale=0.5]{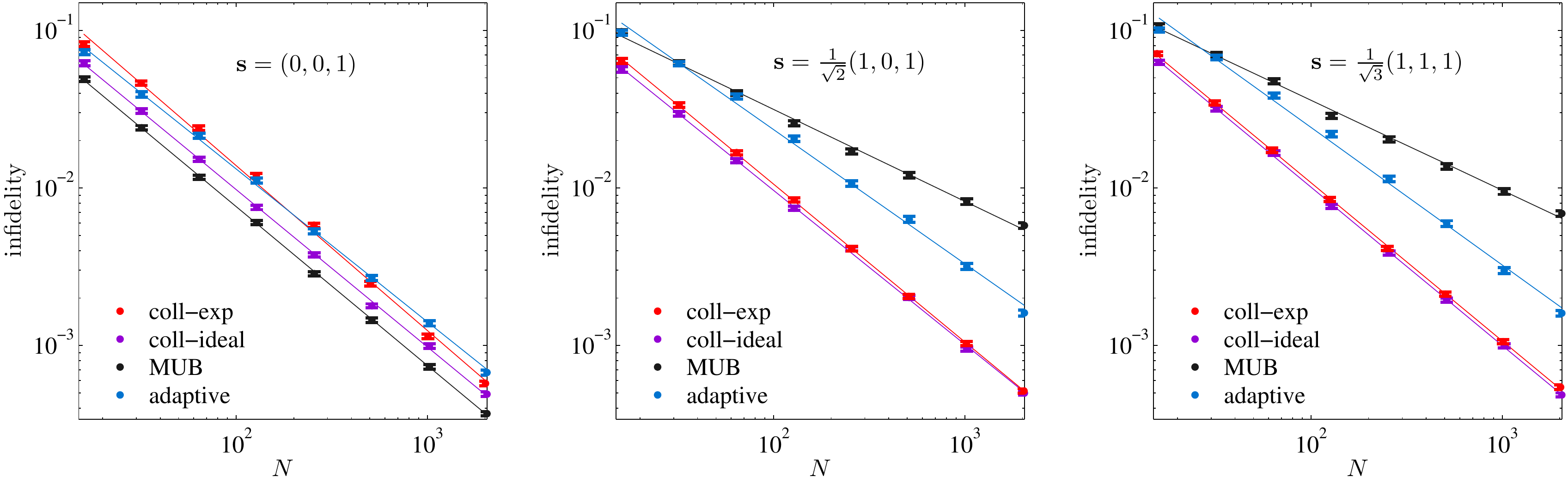}}
	\caption{\label{fig: N} Scaling of the mean infidelity in quantum state tomography with  the collective SIC-POVM (both experiment and simulation). The performances of  MUB  and two-step adaptive measurements  (simulation) are shown for comparison. The three plots  correspond to the tomography of three pure states with Bloch vectors $\vec{s}$ as specified; $N$ is the sample size, ranging from 16 to 2048. Each data point is the average of 1000 repetitions, and the  error bar denotes the standard deviation. }
\end{figure*}

\begin{figure*}
	\center{\includegraphics[scale=0.5]{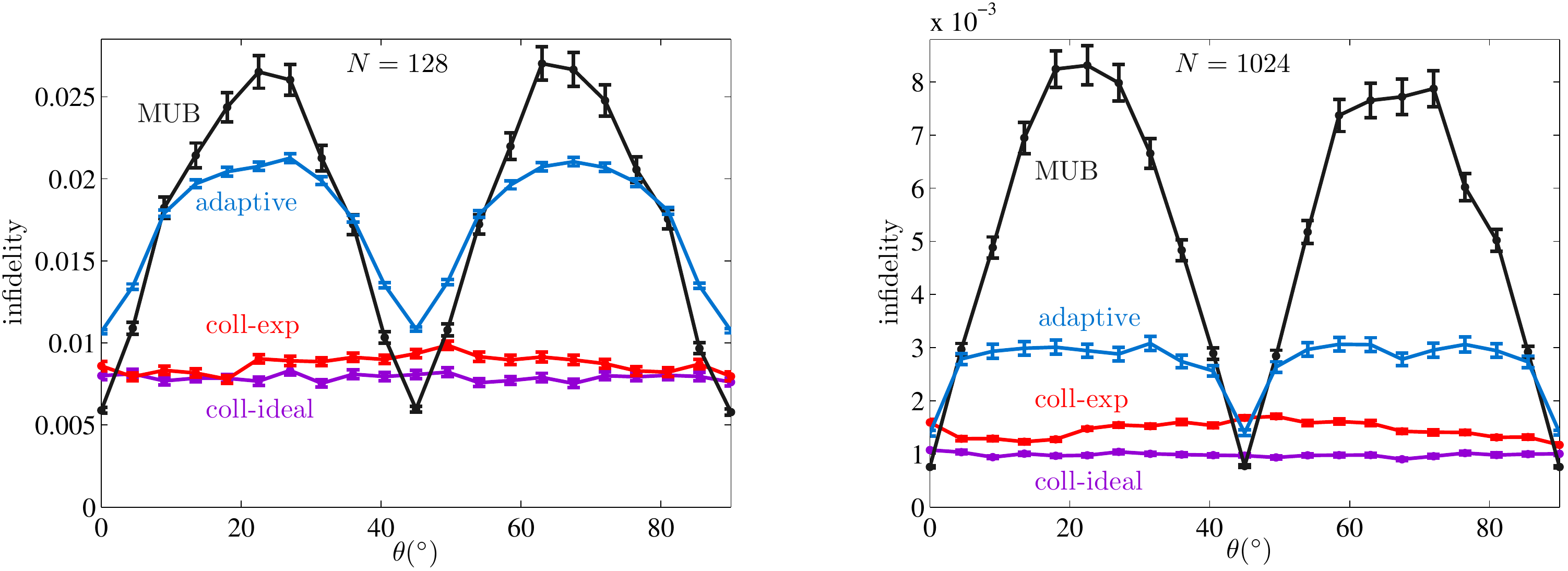}}
	\caption{\label{fig: theta} Mean infidelities achieved by the collective SIC-POVM  in estimating pure states of the form $|\psi(\theta)\>=\sin\theta\ket{0}+\cos\theta\ket{1}$. The performances of  MUB  and two-step adaptive measurements  (simulation) are shown for comparison. 	
		The sample size is  $N=128$ in the left plot and $N=1024$ in the right plot.  Each data point is the average of 1000 repetitions, and the  error bar denotes the standard deviation.}
\end{figure*}

\begin{figure*}
	\center{\includegraphics[scale=0.5]{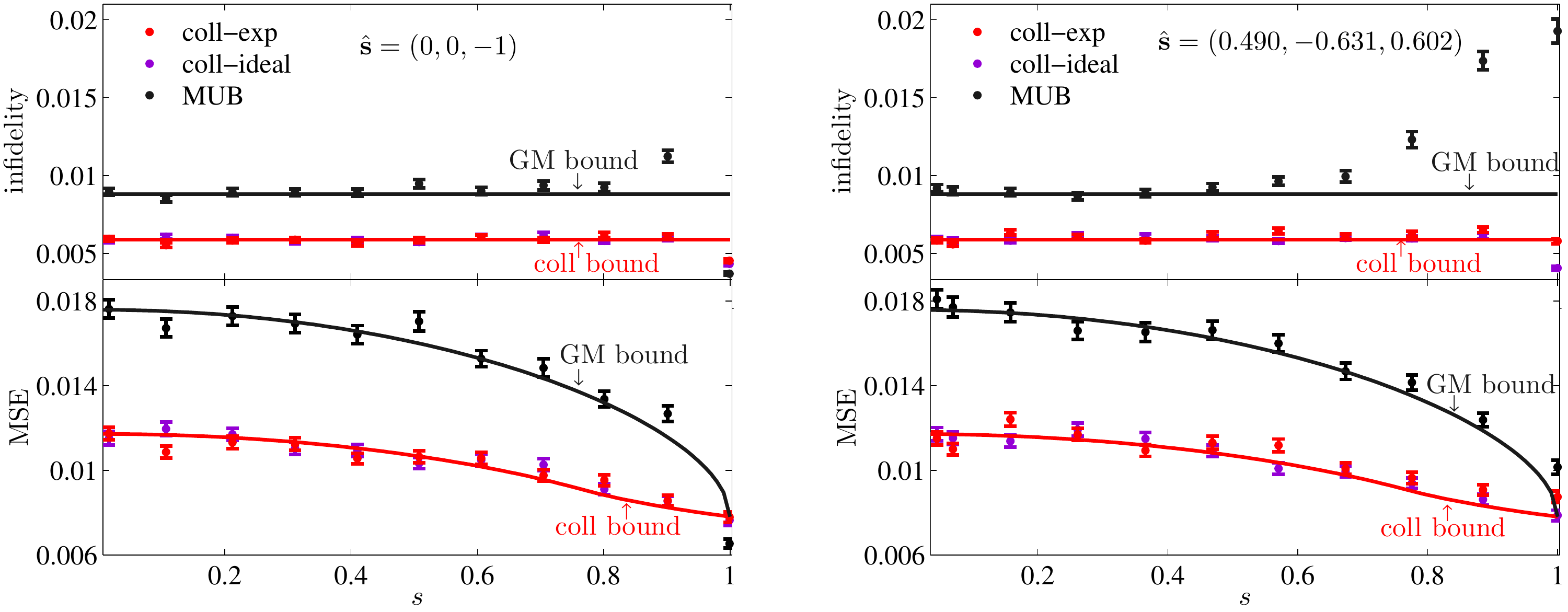}}
	\caption{\label{fig: s} Performance of the collective SIC-POVM in the tomography of mixed qubit states with respect to the mean  infidelity and MSE. Also shown for comparison are the performance of MUB (simulation) as well as  the  Gill-Massar (GM) bound  \cite{GillM00,Zhu12the,HouZXL16} and  a collective (coll) bound \cite{Zhu12the,ZhuH17U} (see the Methods section).
		Here $\hat{\vec{s}}$ and $s$ denote the direction and  length of the Bloch vector; the sample size is $N=256$; each data point is the average of 1000 repetitions, and the  error bar denotes the standard deviation.}
\end{figure*}

In the first experiment, we investigated the scaling of the mean infidelity $1-F$ achieved by the collective SIC-POVM with the sample size $N$ (the number of copies of the state available for tomography).
Three pure states with Bloch vectors $(0,0,1)$, $\frac{1}{\sqrt{2}}(1,0,1)$ and $\frac{1}{\sqrt{3}}(1,1,1)$ were considered (see the supplement for additional results on mixed states). In each case, the probabilities of obtaining the outcomes of the collective SIC-POVM were  estimated from frequencies of repeated measurements, from which we reconstructed  the original state using the maximum likelihood method \cite{PariR04Book}; see the supplement.

 The experimental result and simulation result are shown in   \fref{fig: N}. Also shown as  benchmarks are the simulation results on two popular alternative  schemes: one  based on mutually unbiased bases (MUB) for a qubit \cite{WootF89,DurtEBZ10,AdamS10,Zhu14IOC} and the other based on  two-step adaptive measurements proposed in \rcite{MahlRDF13} (cf.~\rscite{KravSRH13,HouZXL16,qi2017adaptive}).
The experimental result agrees very well with the theoretical predication \cite{ZhuH17U} and numerical simulation. The efficiency of the collective SIC-POVM is almost independent of the input state; the infidelity approximately scales as $O(1/N)$ for all states investigated (cf.~Supplementary \tref{tab:scaling}). This high efficiency  is tied to the fact that the probability of obtaining the outcome $E_5$ in \eref{eq:CollectiveSIC} is very sensitive to the purity of the input state, so that the purity can be estimated very accurately.  By contrast, the scaling behavior is much worse for MUB except when the input  state aligns with one of the POVM elements, which corresponds to "known state tomography"~\cite{MahlRDF13}. This is because the infidelity is very sensitive to  inaccurate estimation of the purity, which is unavoidable for a fixed individual measurement. For a generic pure state, the infidelity achieved by the collective SIC-POVM for $N=2048$ is approximately  twelve (three) times  smaller than that achieved by MUB (local adaptive  measurements). The advantage of the collective SIC-POVM becomes more significant  as the sample size increases.

In the second experiment, we investigated
the mean infidelity achieved by the collective SIC-POVM when the input  states have the form $|\psi(\theta)\>=\sin\theta\ket{0}+\cos\theta\ket{1}$ with  $\theta$ ranging from 0 to $\pi/2$. Here $N$ is chosen to be 128 (that is, 64 pairs) or 1024 (512 pairs).
The result shown in \fref{fig: theta} further demonstrates that  the efficiency of the collective SIC-POVM is almost independent of the input state. In addition, the  infidelity in the worst scenario is much smaller than that achieved by  MUB and local adaptive  measurements. As in the first experiment, the advantage of the collective SIC-POVM becomes more significant when $N$ increases.

In the third experiment, we considered two families of mixed states $\rho=\frac{1}{2}(I+{\vec{s}}\cdot\vec{\sigma})$ with Bloch vectors along $\hat{\vec{s}}=(0,0,-1)$ and $\hat{\vec{s}}=(0.490,-0.631,0.602)$, respectively, and with $s$ ranging from 0 to 1. The sample size  $N$ is chosen to be 256; both the mean infidelity and mean square error (MSE) are considered as figures of merit.  The experimental result is shown in \fref{fig: s}. The mean infidelity achieved by the collective SIC-POVM is not only smaller than that by MUB, but also smaller than the Gill-Massar (GM) bound \cite{GillM00,Zhu12the,HouZXL16}, which constrains the performance of
any local measurement, even with adaptive choices. Moreover, the mean infidelity approximately saturates a bound derived in \rscite{Zhu12the,ZhuH17U}, which represents the best performance that can be achieved by two-copy collective measurements; cf.~the Methods section.
In addition, the collective SIC-POVM   is also nearly optimal with respect to the MSE for all states. Remarkably, such high efficiency is achieved without any adaptive measurement.

\bigskip
\noindent\textbf{Discussion}\\
In summary, we introduced a general method for implementing deterministic collective  measurements on two identically prepared qubits based on quantum walks. Using photonic quantum walks, we then realized experimentally  the collective SIC-POVM  with very  high quality and thereby achieved unprecedented high efficiency in qubit state tomography.
 The collective SIC-POVM we realized  is significantly more efficient than any local measurement. It  improves the scaling of the mean infidelity in the worse scenario from $O(1/\sqrt{N})$ to $O(1/N)$. Moreover,  it is nearly optimal over all two-copy collective measurements with respect to various figures of merit, including the mean infidelity and MSE, although no adaptive measurement is required. This high efficiency manifests the primary advantage of  collective measurements over separable measurements. 
 
 Our work demonstrated a truly nonclassical phenomenon that is owing to entanglement in quantum measurements instead of quantum states.
Moreover, it offers an effective  recipe for exceeding the precision limit of local measurements in quantum state tomography.  
Similar idea can readily  be applied to enhance the precision in multiparameter quantum metrology,  for instance, in the joint estimation of phase and phase diffusion (cf. \rscite{VidrDGJ14,RoccGMS17}), which deserves further study.  
 More generally, our work opens an avenue for exploring the power of collective measurements in quantum information processing. In the future, it would be desirable to extend our approach to realize multi-copy collective measurements on qubits and systems of higher dimensions.

\bigskip

\noindent\textbf{Methods}\\
\noindent \textbf{Coin operators for realizing the collective SIC-POVM.}
Here we present the coin operators that appear in \fref{fig:coll_theory}d; see \sref{sec:QW} in the supplement for more details. 
\begin{align}
&C(-1,1)=\frac{1}{\sqrt{3}}
\begin{pmatrix}
1 & \sqrt{2} \\
\sqrt{2} & -1 \\
\end{pmatrix},  \quad
C(-2,2)=
\begin{pmatrix}
0 & 1 \\
1 & 0 \\
\end{pmatrix}, \nonumber\\
&C(0 ,2)=\frac{1}{2}
\begin{pmatrix}
\sqrt{3} & 1 \\
1 & -\sqrt{3} \\
\end{pmatrix},  \;
C(1,3)=\frac{1}{\sqrt{3}}
\begin{pmatrix}
\sqrt{2} & 1 \\
1 & -\sqrt{2} \\
\end{pmatrix},\nonumber\\
&C(0,4)=
\begin{pmatrix}
1 & 0 \\
0 & -1 \\
\end{pmatrix},\quad
C(-1,5)=\frac{1}{2}
\begin{pmatrix}
1-\rmi & 1+\rmi \\
-1+\rmi & 1+\rmi \\
\end{pmatrix},\nonumber\\
&C(2,2)=C(0,2),\quad  C(-1,3)=C(-1,1), \nonumber\\ &C(-2,4)=C(-2,2).
\end{align}

\bigskip
\noindent \textbf{Fidelity between two POVMs.}
Consider two POVMs $\{E_j\}_{j=1}^M$ and  $\{E_j^\prime\}_{j=1}^M$ on a $d$-dimensional Hilbert space with the same number of elements, where $E_j^\prime$ is the counterpart of $E_j$ (for example, one is the experimental realization of the other). Construct two normalized quantum states as $\sigma=\frac{1}{d}\sum_{j=1}^M E_j\otimes(\outer{j}{j})$ and $\sigma^\prime=\frac{1}{d}\sum_{j=1}^M E_j^\prime\otimes(\outer{j}{j})$, where $|j\>$ form an orthonormal basis for an ancilla system. The fidelity between the  two POVMs $\{E_j\}_{j=1}^M$ and $\{E_j^\prime\}_{j=1}^M$  is defined as the fidelity between the two states $\sigma$ and $\sigma'$,
\begin{equation}\label{eq:Fidelity}
F(\sigma,\sigma^\prime):=\Bigl(\tr \sqrt{\sqrt{\sigma}\sigma^\prime\sqrt{\sigma}}\Bigl)^2=\biggl(\sum_{j=1}^M w_j \sqrt{F_j}\biggr)^2,
\end{equation}
where $w_j=\frac{\sqrt{\tr (E_j)\tr (E_j^\prime)}}{d}$, and $F_j=F\Bigl(\frac{E_j}{\tr (E_j)},\frac{E_j^\prime}{\tr (E_j^\prime)}\Bigr)$ is the fidelity between the two normalized POVM elements $\frac{E_j}{\tr (E_j)}$ and $\frac{E_j^\prime}{\tr (E_j^\prime)}$.

\bigskip
\noindent \textbf{Gill-Massar bounds and collective bounds.}
In quantum state tomography with individual measurements
(including local adaptive measurements), the precision achievable is  constrained by the Gill-Massar (GM) bound \cite{GillM00,Zhu12the,HouZXL16} (see also \rcite{Haya97}). In the case of a qubit, the GM bound is $\frac{9}{4N}$ when the figure of merit is the mean infidelity (approximately equal to the mean square Bures distance), where $N$ is the sample size (assuming $N$ is not too small). When the figure of merit is the MSE $\mathbb{E}\tr[(\hat{\rho}-\rho)^2]$, the GM bound is  $\frac{(2+\sqrt{1-s^2})^2}{2N}$, where $s$ is the length of the Bloch vector of the qubit state.

When  collective measurements on two identical qubits are allowed, the precision limit is constrained by a collective bound. According to Eqs.~(6.73) and (6.74)  in \rcite{Zhu12the} with $t=3/2$,
the collective bound
for the mean infidelity (mean square Bures distance) is $\frac{3}{2N}$ (cf.~\rcite{ZhuH17U}), and the bound for
the MSE is
\begin{equation}
\begin{cases}
\frac{(2+\sqrt{1-s^2})^2}{3N} &\mbox{if}\quad 0\leq  s\leq \frac{3+4\sqrt{3}}{13},\\
\frac{ s(1+s)(3-s)}{(3s-1)N} &\mbox{if}\quad \frac{3+4\sqrt{3}}{13}\leq s\leq 1.
\end{cases}
\end{equation}

The GM bound and collective bound for the mean infidelity may be violated when the state is nearly pure (with thresholds depending on $N$), in which case common estimators (including the maximum likelihood estimator) are  biased due to  the boundary of the state space.  The precision limits with respect to the MSE are less sensitive to this influence.

\noindent \textbf{Data availability.} The data that support the results of this study are available from the corresponding author upon request.

\bigskip

\bigskip
\noindent \textbf{Acknowledgements}\\
The work at USTC is supported by the National Natural Science Foundation of China under Grants (Nos. 11574291, 11774334, 61327901 and 11774335),  the National Key Research and Development Program of China (No.2017YFA0304100),  Key Research Program of Frontier Sciences, CAS (No.QYZDY-SSW-SLH003), the Fundamental Research Funds for the Central Universities (No.WK2470000026) and  China Postdoctoral Science Foundation (Grant No.2016M602012). HZ acknowledges financial support from the Excellence Initiative of the German Federal and State Governments (ZUK~81) and the DFG. JS acknowledges financial support from the ERC (Consolidator Grant 683107/TempoQ), and the DFG.

\bigskip
\noindent \textbf{Author contributions}\\
HZ developed the theoretical approach; GYX supervised the project; ZBH, JFT, JL and GYX designed the experiment and the measurement apparatus for the collective measurement; ZBH built the instruments, performed the experiment and collected the data with assistance from GYX, JFT, YY and KDW; JS developed the maximum likelihood algorithm for state tomography with collective measurements. ZBH, JS, HZ and GYX performed numerical simulations and analysed the experimental data with assistance from CFL and GCG; HZ, ZBH, JS and GYX prepared and wrote the manuscript.

\bigskip
\noindent \textbf{Competing financial interests}\\
The authors declare no competing financial interests.

 \clearpage
 \newpage
 \addtolength{\textwidth}{-1in}
 \addtolength{\oddsidemargin}{0.5in}
 \addtolength{\evensidemargin}{0.5in}

 \setcounter{equation}{0}
 \setcounter{figure}{0}
 \setcounter{table}{0}

 \makeatletter
 \renewcommand{\theequation}{S\arabic{equation}}
 \renewcommand{\thefigure}{S\arabic{figure}}
 \renewcommand{\thetable}{S\arabic{table}}


 \onecolumngrid
 \begin{center}
 	\textbf{\large  Deterministic realization of efficient collective measurements via photonic quantum walks: Supplement}
 \end{center}

\section{\label{sec:QW}Realization of the collective SIC-POVM via quantum walks}
Recently, quantum walks were proposed as a recipe for implementing general POVMs on a single qubit \cite{KurzW13}, which have been demonstrated in experiments \cite{BianLQZ15,ZhaoYKX15}. In a one-dimensional discrete quantum walk, the system state is characterized by two degrees of freedom $|x,c\>$, where $x=\cdots, -1,0,1,\cdots$ denotes the walker position, and $c=0,1$ represents the coin state. The dynamics of each step is described by a unitary transformation of the form $U(t)=TC(t)$, where $T$ is the conditional translation operator
\begin{align}
T = \sum_x \ket{x+1,0}\bra{x,0} + \ket{x-1,1}\bra{x,1},
\end{align}
and $C(t)=\sum_x |x\>\<x|\otimes C(x,t)$  with $C(x,t)$ being site-dependent coin operators. A general POVM on a qubit can be realized by engineering the coin operators $ C(x,t)$ followed by measuring the walker position after certain steps. However, little is known in the literature on realizing POVMs on higher dimensional systems based on  quantum walks. Here we offer a recipe to extending the capabilities of quantum walks.

For concreteness, we focus on the collective SIC-POVM  on a two-qubit system, which is composed of five outcomes,
\begin{equation}\label{eq:CollectiveSICsupp}
E_j =\frac{3}{4}(|\psi_j\>\<\psi_j|)^{\otimes 2}, \quad E_5=|\Psi_-\>\<\Psi_-|,
\end{equation}
where $|\Psi_-\>=\frac{1}{\sqrt{2}}(|\zero\one\>-|\one\zero\>)$ is the singlet, and
\begin{equation}\label{eq:QubitSICsupp} 
\begin{aligned}
\ket{\psi_1}=&\ket{\zero},& \ket{\psi_2}=&\frac{1}{\sqrt{3}}(\ket{\zero}+\sqrt{2}\ket{\one}),\\ \ket{\psi_3}=&\frac{1}{\sqrt{3}}(\ket{\zero}+\rme^{\frac{2\pi}{3}\rmi}\sqrt{2}\ket{\one}),\quad & \ket{\psi_4}=&\frac{1}{\sqrt{3}}(\ket{\zero}+\rme^{-\frac{2\pi}{3}\rmi}\sqrt{2}\ket{\one})
\end{aligned}
\end{equation}
form a
symmetric informationally complete POVM (SIC-POVM) on a qubit \cite{Zaun11,ReneBSC04}. Here the "overline" on $\zero, \one$ is added to distinguish logical quantum states from physical quantum states of the walker and  the coin.
The  Bloch vectors of the four states $|\psi_j\>$ for $j=1,2,3,4$  are given by
$\vec{r}_1=(0, 0, 1)$, $\vec{r}_2=(\frac{2\sqrt{2}}{3}, 0, -\frac{1}{3})$, $\vec{r}_3=(-\frac{\sqrt{2}}{3}, \frac{\sqrt{6}}{3}, -\frac{1}{3})$ and $\vec{r}_4=(-\frac{\sqrt{2}}{3}, -\frac{\sqrt{6}}{3}, -\frac{1}{3})$, which form a regular tetrahedron inside the Bloch sphere.

To realize the collective SIC-POVM using quantum walks, the coin qubit and the walker in positions~1 and $-1$ are taken as the two-qubit system of interest, while the other positions of the walker act as an ancilla. With this choice,
the collective SIC-POVM can be realized via five-step quantum walks as illustrated in \fref{fig:coll_theory}d in the main text, with  nontrivial coin operators given by
\begin{equation}\label{eq:coinOperatorColl}
\begin{aligned}
C(-1,1)=&\frac{1}{\sqrt{3}}\left(
 \begin{array}{cc}
 1 & \sqrt{2} \\
 \sqrt{2} & -1 \\
 \end{array}
 \right), \quad &
 C(-2,2)=&\left(
 \begin{array}{cc}
 0 & 1 \\
 1 & 0 \\
 \end{array}
 \right), \quad &
 C(0 ,2)=&\frac{1}{2}\left(
 \begin{array}{cc}
 \sqrt{3} & 1 \\
 1 & -\sqrt{3} \\
 \end{array}
 \right), \\
  C(1,3)=&\frac{1}{\sqrt{3}}\left(
 \begin{array}{cc}
 \sqrt{2} & 1 \\
 1 & -\sqrt{2} \\
 \end{array}
 \right), &
 C(0,4)=&\left(
 \begin{array}{cc}
 1 & 0 \\
 0 & -1 \\
 \end{array}
 \right),&
 C(-1,5)=&\frac{1}{2}\left(
 \begin{array}{cc}
 1-\rmi & 1+\rmi \\
 -1+\rmi & 1+\rmi \\
 \end{array}
 \right),\\
 C(2,2)=&C(0,2),&  C(-1,3)=&C(-1,1), & C(-2,4)=&C(-2,2).
\end{aligned}
\end{equation}
To see this, note that a general logical two-qubit pure state
\begin{equation}
\ket{\Psi_{0}} = a\ket{\zero\zero} + b\ket{\zero\one} + c\ket{\one\zero} + d\ket{\one\one},\quad  |a|^2 + |b|^2 + |c|^2 + |d|^2 = 1
\end{equation}
can be encoded into the initial  state (corresponding to step $t=0$)  of the walker-coin system as
\begin{equation}
\ket{\Psi_{0}}= a\ket{1,0} + b\ket{1,1} + c\ket{-1,0} + d\ket{-1,1}.
\end{equation}
After step~1, the state $\ket{\Psi_{0}}$ evolves into
\begin{equation}
\ket{\Psi_1} = TC(t=1)\ket{\Psi_0}=a\ket{2,0} + b\ket{0,1} + \biggl(\sqrt{\frac{1}{ 3}}\,c + \sqrt{\frac{2}{ 3}}\,d\biggr)\ket{0,0} + \biggl(\sqrt{\frac{2}{ 3}}\,c - \sqrt{\frac{1}{ 3}}\,d\biggr)\ket{-2,1}.
\end{equation}
Following a similar procedure, the state after step~5 reads
\begin{align}
\ket{\Psi_5} =&\frac{\sqrt{3} }{ 2}a\ket{6,0} +
\frac{\sqrt{3}}{ 6}(a +\sqrt{2} b + \sqrt{2}c + 2d)\ket{4,0} +\frac{\sqrt{2}}{ 2}(-b + c)\ket{2,0}\nonumber\\
&+\frac{\sqrt{3}}{ 6}\rme^{\frac{\pi}{4}\rmi}\big( a + {\sqrt{2}}\rme^{-{\frac{2\pi}{3} \rmi}}b + {\sqrt{2}}\rme^{-{\frac{2\pi}{3} \rmi}}c +{2 }\rme^{-{\frac{4\pi}{3} \rmi}}d\big)\ket{0,0}\nonumber\\
& + \frac{\sqrt{3}}{ 6}\rme^{\frac{\pi}{4}\rmi}\big( a + {\sqrt{2}}\rme^{{\frac{2\pi}{3} \rmi}}b + {\sqrt{2} }\rme^{{\frac{2\pi}{3} \rmi}}c + {2 }\rme^{{\frac{4\pi}{3} \rmi}}d\big)\ket{-2,1}.
\end{align}
Now measuring the position of the walker realizes the collective SIC-POVM as desired. To verify this claim, note that the probabilities of detecting the walker at positions $6, 4, 2,0, -2$ are respectively given by
\begin{equation}
\begin{aligned}
\tilde{p}_6=& \frac{3}{4}|a|^2=\frac{3}{4}|\<\psi_1\psi_1|\Psi_0\>|^2=\<\Psi_0|E_1|\Psi_0\>,\\
\tilde{p}_4=& \frac{1}{12}|a +\sqrt{2} b + \sqrt{2}c + 2d|^2=\frac{3}{4}|\<\psi_2\psi_2|\Psi_0\>|^2=\<\Psi_0|E_2|\Psi_0\>, \\
\tilde{p}_2=& \frac{1}{2}|-b + c|^2=|\<\Psi_-|\Psi_0\>|^2=\<\Psi_0|E_5|\Psi_0\>, \\
\tilde{p}_0=& \frac{1}{12}\bigl| a + {\sqrt{2}}\rme^{-{\frac{2\pi}{3} \rmi}}b + {\sqrt{2}}\rme^{-{\frac{2\pi}{3} \rmi}}c +{2 }\rme^{-{\frac{4\pi}{3} \rmi}}d\bigr|^2=\frac{3}{4}|\<\psi_3\psi_3|\Psi_0\>|^2=\<\Psi_0|E_3|\Psi_0\>,
\\
\tilde{p}_{-2}=& \frac{1}{12}\bigl| a + {\sqrt{2}}\rme^{{\frac{2\pi}{3} \rmi}}b + {\sqrt{2}}\rme^{{\frac{2\pi}{3} \rmi}}c +{2 }\rme^{{\frac{4\pi}{3} \rmi}}d\bigr|^2=\frac{3}{4}|\<\psi_4\psi_4|\Psi_0\>|^2=\<\Psi_0|E_4|\Psi_0\>,
\end{aligned}
\end{equation}
So  the detectors at the five positions  $6, 4, 2,0, -2$ correspond to the five POVM elements $E_1, E_2, E_5, E_3, E_4$ specified in  \esref{eq:CollectiveSICsupp} and \eqref{eq:QubitSICsupp}.
Note that a detector  at position 6 after  step~5 is equivalent to a detector at position 3 after step~2. Similarly, the detector at 
position~4 (position~2) after step~5  can be placed at
position~2 (position~1) after step~3 (step~4) without changing the detection probability. This fact can be utilized to simplify the experimental design, as reflected in  \fref{fig:coll_theory} in the main text.

\begin{figure*}
	\center{\includegraphics[scale=0.62]{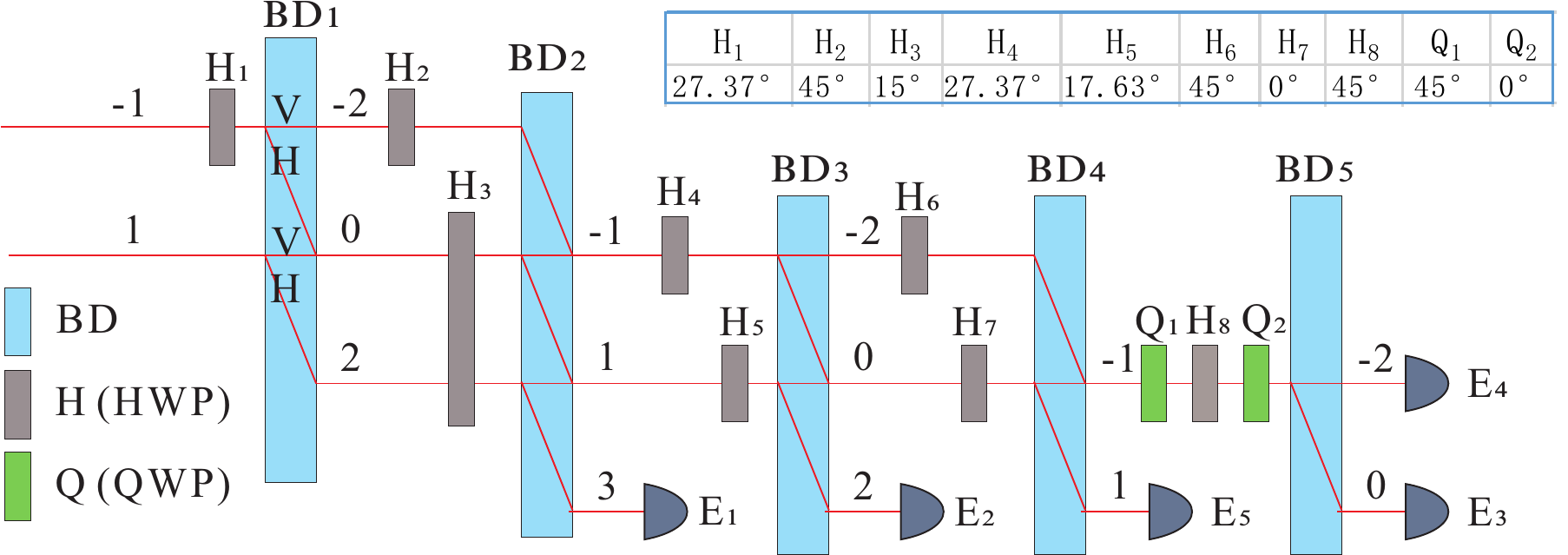}}
	\caption{\label{fig:coll_theorySupp}
		Realization of the collective SIC-POVM using five-step photonic quantum walks (cf.~\fref{fig:coll_theory}d and \fref{fig: exp setup}  in the main text). The polarization-encoded coin qubit and the walker in positions 1 and $-1$ are taken as the two-qubit system of interest, while the other positions of the walker act as an ancilla. Beam displacers  (BDs) are used to realize the conditional translation operator $T$.  Combinations of half wave plates (HWPs) and quarter wave plates (QWPs) are used to realize site-dependent  coin operators $C(x,t)$. Five single-photon detectors (SPDs) $E_1$ to $E_5$  correspond to the five outcomes. The collective SIC-POVM is realized by choosing the rotation angles  of the HWPs and QWPs according to the table embedded in the figure (identical to the table embedded in \fref{fig: exp setup}). 				
	}
\end{figure*}

The above proposal can be realized using photonic quantum walks as illustrated in  \fref{fig:coll_theorySupp} as well as \fref{fig: exp setup} in the main text.  In this scheme, the conditional translation operator is realized by  beam displacers  (BDs), which displace the H-component  away from the V-component. The coin operators are realized by suitable combinations of  half wave plates (HWPs) and quarter wave plates (QWPs)
with rotation angles specified in \fref{fig:coll_theorySupp}.
Note that a HWP with rotation angle $h$ and a QWP with  rotation angle $q$ realize the following  unitary transformations
 \begin{equation}
 U_{\mathrm{H}}(h) = \left(
 \begin{array}{cc}
 \cos2h & \sin2h \\
 \sin2h & -\cos2h \\
 \end{array}
 \right),\quad U_{\mathrm{Q}}(q) =  \frac{1}{\sqrt{2}}\rme^{\frac{\pi}{4}\rmi}\left(
 \begin{array}{cc}
 1-\rmi\cos2q & -\rmi\sin2q \\
 -\rmi\sin2q & 1+\rmi\cos2q \\
 \end{array}
 \right).
 \end{equation}
 Based on this equation, it is straightforward to verify that the site-dependent coin operators presented in \eref{eq:coinOperatorColl} are realized by the wave plates shown in \fref{fig:coll_theorySupp} with rotation angles as specified.

\section{\label{sec:StatePre}State preparation}

In this section we provide additional details on the preparation of  walker-coin two-qubit states considered in the main text. Our discussion is based on the
 state preparation module illustrated in  \fref{fig: exp setup}. Note that the quartz crystal and the HWP corresponding to $\alpha_2$ are used only in the  preparation of mixed states. In general, a walker-coin two-qubit state can be prepared by properly choosing the rotation angles $\alpha_1$,$\alpha_2$, $h_1$, $h_2$, $h_3$, $q_1$ and $q_2$ of the HWPs and QWPs shown in this module. The parameter choices for various  states considered in the main text are specified in \tref{state_preparation_angles}.
 Here $\ket{\pm_z}$, $\ket{\pm_x}$ and $\ket{\pm_y}$ denote the two eigenstates with eigenvalues $\pm1$ of $\sigma_z$, $\sigma_x$ and  $\sigma_y$, respectively, with Bloch vectors given by $(0,0,\pm1)$, $(\pm1,0,0)$ and $(0,\pm1,0)$; these states are used in the measurement tomography of the collective SIC-POVM. $\hat{E}_j$ for $j=1$ to 5 denote the five normalized POVM elements of the collective SIC-POVM (note that $\hat{E}_1=\ket{+_z+_z}\bra{+_z+_z}$); these states are used in the experimental verification of the  collective SIC-POVM. The rest states in the table are studied in quantum state tomography with the collective SIC-POVM. To be specific,
 $\ket{\psi(\theta)}=\sin\theta\ket{0}+\cos\theta\ket{1}$ is a  pure state parametrized by $\theta$;  $\frac{1}{\sqrt{2}}(1,0,1)$ and  $\frac{1}{\sqrt{3}}(1,1,1)$ denote two pure  states with Bloch vectors as specified;  $\vec{s}_1$ and $\vec{s}_2$ denote quantum states whose Bloch vectors align with
 $\hat{\vec{s}}_1=(0,0,-1)$ and $\hat{\vec{s}}_2=(0.490,-0.631,0.602)$, and with lengths $s_1$ and $s_2$, respectively.

In the preparation of $\hat{E}_5$, which is maximally entangled, the QWP corresponding to $q_2$ is removed, and $h_3$ is set at  $0^\circ$.  For all other states considered in this work, which are product states, this QWP is present, and
 $h_3$ is set at  $45^\circ$.
In the preparation of a product state, $\alpha_1$, $h_1$, $q_1$ are used to control the state of the first qubit (walker), while $\alpha_2$ (together with the quartz crystal), $h_2$, $q_2$ are used to control the state of the second qubit (coin).
Specifically, the length of the Bloch-vector  of the first qubit is determined by $\alpha_1$, while the direction of the Bloch-vector is determined by $h_1$, $q_1$.
In the preparation of the second qubit, $\alpha_2$, $h_2$, $q_2$ play similar roles to $\alpha_1$, $h_1$, $q_1$ for the first qubit. The parameters shown in \tref{state_preparation_angles} apply to the preparation of a two-copy state,   of which the walker qubit and the coin qubit are identical.  Since the preparation of the two qubit states are independent,  product states with different marginals can also be prepared with straightforward modification. For example, the product state $\ket{+_z+_x}$ can be prepared by choosing  the following parameters:
\begin{equation}
\alpha_1=q_1=0, \quad h_1=h_3=q_2=45^\circ,\quad h_2=-22.5^\circ.
\end{equation}

\begin{table}
	\caption{\label{state_preparation_angles}Parameter choices in walker-coin two-qubit state preparation. Here
	$\alpha_1, \alpha_2, h_1, h_2, q_1, q_2$ are the rotation angles of the HWPs and QWPs shown in \fref{fig: exp setup}, and "quartz" denotes the quartz crystal in the same figure. 	
	 $\ket{\pm_z}$, $\ket{\pm_x}$ and $\ket{\pm_y}$ denote the two eigenstates with eigenvalues $\pm1$ of $\sigma_z$, $\sigma_x$ and  $\sigma_y$; the parameters in the parentheses apply to the states with eigenvalue $-1$. $\hat{E}_j$ for $j=1$ to 5 denote the five normalized POVM elements of the collective SIC-POVM (note that $\hat{E}_1=\ket{+_z+_z}\bra{+_z+_z}$).
	$\ket{\psi(\theta)}=\sin\theta\ket{0}+\cos\theta\ket{1}$ is a  pure state.  $\frac{1}{\sqrt{2}}(1,0,1)$ and  $\frac{1}{\sqrt{3}}(1,1,1)$ denote two pure  states with Bloch vectors as specified.  $\vec{s}_1$ and $\vec{s}_2$ denote quantum states whose Bloch vectors  align with
	$\hat{\vec{s}}_1=(0,0,-1)$ and $\hat{\vec{s}}_2=(0.490,-0.631,0.602)$, and with lengths $s_1$ and $s_2$, respectively; $\alpha(s)=\frac{1}{4}\arccos(s)$.  The quartz and the HWP corresponding to $\alpha_2$ are used only in the  preparation of mixed states (in the last two columns of the table). In the preparation of $\hat{E}_5$, the QWP corresponding to $q_2$ is removed. 
	}
	\begin{tabular}{c c c c c c c c c c c c c c}
		\hline\hline
		States & $\ket{\pm_z}$ & $\ket{\pm_x}$ & $\ket{\pm_y}$ & $\hat{E}_2$& $\hat{E}_3$ & $\hat{E}_4$& $\hat{E}_5$ & $\frac{1}{\sqrt{2}}(1,0,1)$& $\frac{1}{\sqrt{3}}(1,1,1)$&$\ket{\psi(\theta)}$& $\vec{s}_1$& $\vec{s}_2$\\
		\hline
		$\alpha_1$ &   0 & 0  & 0   &  0& 0 & 0  &  0& 0 & 0  &0&  $\alpha(s_1)$&  $\alpha(s_2)$\\
		$h_1(^\circ)$& 45(0)  & -22.5(22.5) &-22.5(22.5)  &  -17.63& 0 & 27.37  &  22.5& 56.25 & -24.95& $90-\frac{\theta}{2} $& 0 & 45  \\
		$q_1(^\circ)$& 0 & 45  & 0   &  -35.26 & 27.37 & 27.37  &  45& 22.5 & 22.5  &$-\theta$&  0&  19.57\\
		$h_3(^\circ)$& 45 & 45 &45  &  45& 45 &45  &  0& 45 & 45  &  45&  45&  45 \\
		$\alpha_2(^\circ)$ & $-$ & $-$     &  $-$  & $-$  & $-$  &  $-$  & $-$  & $-$  &  $-$&  $-$  &  $\alpha(s_1)$&  $\alpha(s_2)$\\
		quartz & $-$  & $-$   &  $-$   & $-$  & $-$  & $-$   & $-$  & $-$  & $-$   &  $-$&  0&  0\\
		$h_2(^\circ)$& 45(0) & -22.5(22.5) & -22.5(22.5)  &  -17.63& 0 & 27.37  &  45& 56.25 & -24.95 & $90-\frac{\theta}{2}$ &  0&  45 \\
		$q_2(^\circ)$& 0& 45  & 0   &  -35.26 & 27.37 & 27.37  &   $-$& 22.5 & 22.5 &$-\theta$ &  0&  19.57\\
		\hline\hline
	\end{tabular}
\end{table}
\bigskip

\section{\label{sec:MeasTomo}Quantum measurement tomography of the collective SIC-POVM}

\begin{figure*}
	\center{\includegraphics[scale=0.42]{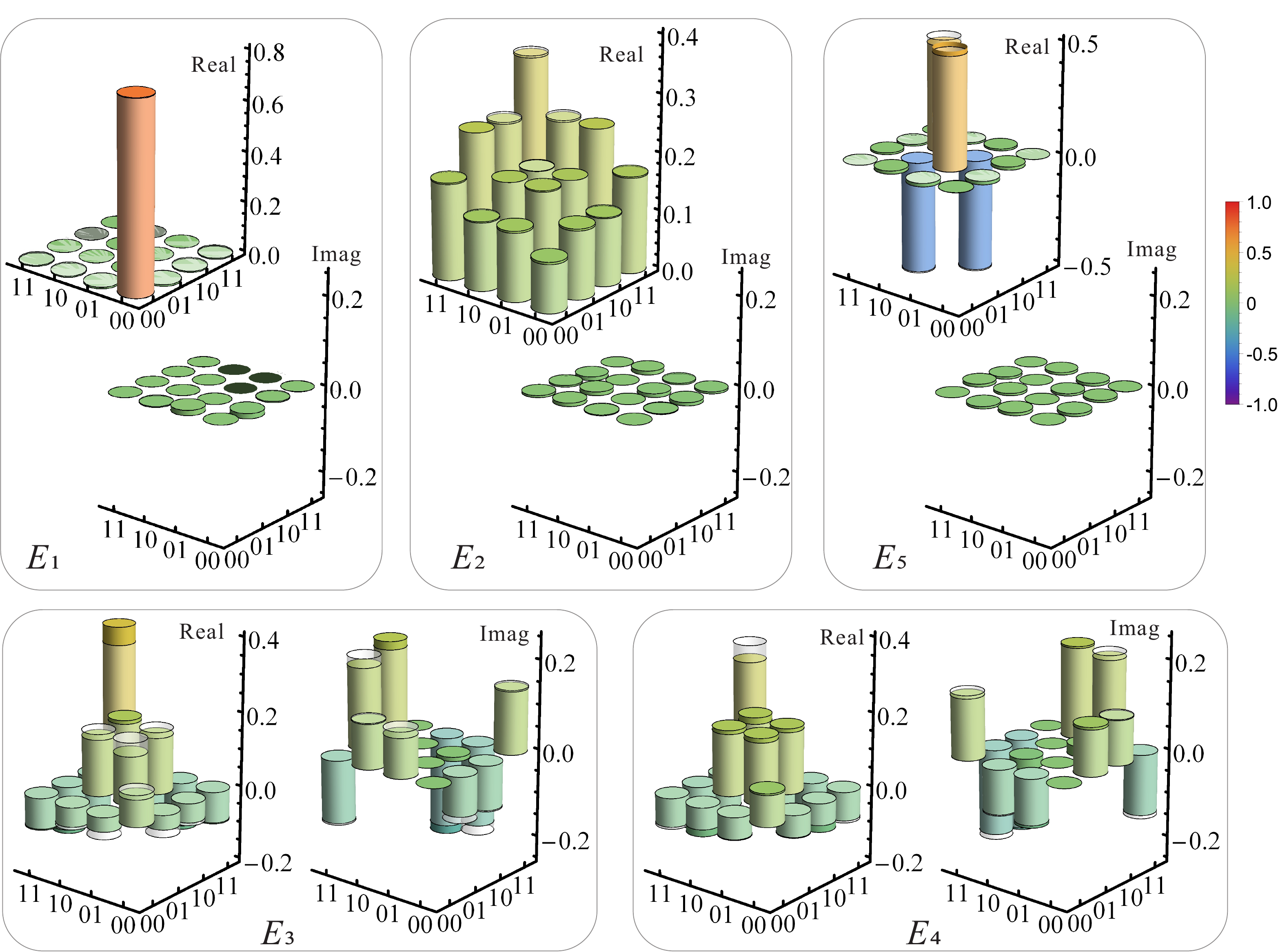}}
	\caption{\label{fig:densitymaxtrix} Results on measurement tomography of the collective SIC-POVM realized in the experiment. The matrix elements of the real (Real) and imaginary (Imag) parts of the five POVM elements $E_1$ to $E_5$ are plotted using solid colours. For comparison, the counterparts of the ideal POVM are plotted as wire frames.		}
\end{figure*}

In this section, we provide more details on the measurement tomography of the collective SIC-POVM realized using photonic quantum walks. To perform measurement tomography, 36 states,  the tensor products of the six eigenstates of three Pauli operators, were prepared according to the method described in the previous section
and sent to the collective-measurement module. To reduce statistical fluctuation, each state was prepared and measured
35000 times. Then the five POVM elements were estimated from the  measurement statistics using the maximum likelihood (ML) method developed in \rcite{Fiur01maximum}. The five reconstructed POVM elements are shown in \fref{fig:densitymaxtrix} in comparison with the ideal counterparts. The fidelities of the five  POVM elements   are $0.9991\pm0.0001$,   $0.9979\pm0.0007$,        $0.9870\pm0.0008$,     $0.9927\pm0.0002$ and  $0.9961\pm0.0002$, respectively;  the overall fidelity of the POVM is
$0.9946\pm0.0002$  (cf.~the Methods section).  These results show that the collective SIC-POVM was realized with very high quality.

\section{Scaling  of the mean infidelity with the sample size}
In this supplement, we provide additional details on the scaling of the mean infidelity with the sample size in quantum state tomography with the collective SIC-POVM. To complement the results presented in the main text, we first illustrate the scaling of the mean infidelity $1-F$ achieved by the collective SIC-POVM for mixed states. We then present the scaling exponents for both pure states and mixed states.

\subsection{\label{sec:ScalingMix}Scaling  of the mean infidelity  for   mixed states}
\begin{figure*}
	\center{\includegraphics[scale=0.47]{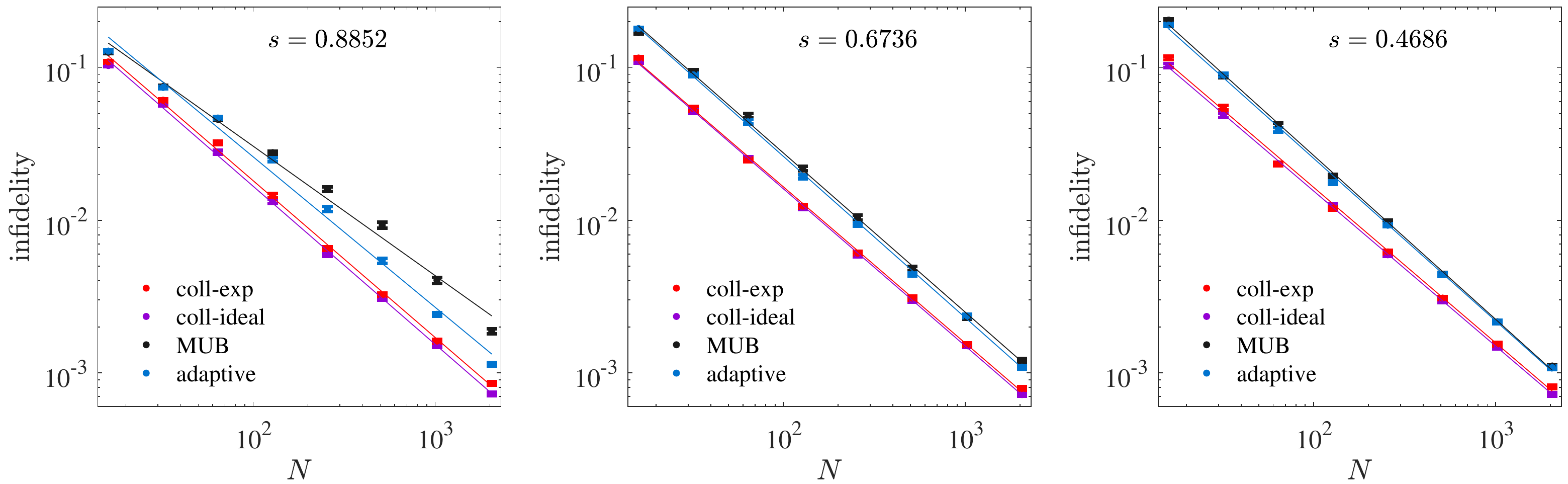}}
	\caption{ \label{fig: N mixed} Scaling of the mean infidelity in the tomography of  mixed states with  the collective SIC-POVM (both experiment and simulation). The performances of  MUB and two-step adaptive measurements (simulation) are shown for comparison. The three plots  correspond to the tomography of three mixed states whose  Bloch vectors align with the same direction $\hat{\vec{s}}=(0.490,-0.631,0.602)$ and have  lengths $s=0.885$, $0.674$ and $0.469$, respectively. Here $N$ is the sample size, ranging from 16 to 2048. Each data point is the average of 1000 repetitions, and the  error bar denotes the standard deviation. }
\end{figure*}

Three mixed states were considered; their Bloch vectors align with the same randomly-chosen direction
 $\hat{\vec{s}}=(0.490,-0.631,0.602)$, with  lengths $s=0.885$, $0.674$ and $0.469$, respectively. The experimental result as well as the simulation result on the ideal collective SIC-POVM are shown in   \fref{fig: N mixed}. Also shown as  benchmarks are the simulation results on the performances of two popular alternative  schemes based on mutually unbiased bases (MUB) for a qubit \cite{WootF89,DurtEBZ10,AdamS10,Zhu14IOC}.
Similar to the case of pure states, the efficiency of the collective SIC-POVM  is almost independent of the input state; 
the infidelity approximately scales as $O(1/N)$ for all states investigated.

By contrast, the scaling behavior for MUB is sensitive to the purity of the input state. 
When the input mixed state has a high purity (see the left plot in \fref{fig: N mixed}), the infidelity  scales as $O(1/\sqrt{N})$ when $N$ is small, while it scales as $O(1/N)$ when $N$ is large. The transition region  depends on the purity of the input state. In the special case of  a pure state ($s=1$),
  the $O(1/\sqrt{N})$ scaling approximately holds for all $N$ (see the middle and right plots in \fref{fig: N}). When the state is highly mixed, MUB achieves almost the same scaling
   $O(1/N)$ as the collective SIC-POVM, but the infidelity is still larger by a constant factor of about 1.5   (see the right plot in \fref{fig: N mixed}).

\subsection{Scaling  exponents}
In the main text, we investigated the scaling of the mean infidelity $1-F$ achieved by the collective SIC-POVM with the sample size $N$.
Three pure states with Bloch vectors $(0,0,1)$, $\frac{1}{\sqrt{2}}(1,0,1)$ and $\frac{1}{\sqrt{3}}(1,1,1)$ 
were considered. The experimental result as well as the simulation result on the ideal collective SIC-POVM are shown in   \fref{fig: N} in the main text. Also shown as  benchmarks are the simulation results on the performances of two popular alternative  schemes: one  based on mutually unbiased bases (MUB) for a qubit \cite{WootF89,DurtEBZ10,AdamS10,Zhu14IOC} and the other based on  two-step adaptive measurements proposed in \rcite{MahlRDF13} (cf.~\rscite{KravSRH13,HouZXL16,qi2017adaptive}). In addition, three mixed states were investigated in \sref{sec:ScalingMix} in this supplement; see \fref{fig: N mixed}.

To quantify the distinction between different measurement schemes,
experimental data and numerical data  are fitted to power laws of the form  $1-F=\beta N^{-p}$. The exponents~$p$ are shown in \tref{tab:scaling}.
According to this table, the efficiency of the collective SIC-POVM is almost independent of the input state; the infidelity approximately scales as $O(1/N)$ for all states investigated. By contrast, the scaling behavior  for MUB is very sensitive to the input state. For states with high purities, the scaling is usually  much worse except when the input  state aligns with one of the POVM elements, which corresponds to "known state tomography" \cite{MahlRDF13}.

\begin{table}
	\caption{\label{tab:scaling} Scaling exponents of the  mean infidelity $1-F$ against the sample size $N$ in quantum state tomography.   The performance of the  collective SIC-POVM (both experiment and simulation) is compared with that of MUB and local adaptive measurements (simulation). The first three columns represent the results on three pure states with Bloch vectors as specified, while the last three columns represent results on three mixed states whose Bloch vectors align with the same direction $\hat{\vec{s}}=(0.490,-0.631,0.602)$, with  lengths $s$ as specified. The data presented in  \fsref{fig: N} and \ref{fig: N mixed} are
		fitted to the formula $1-F=\beta N^{-p}$. The scaling exponents $p$ for the above six states and  four measurement schemes are shown in this table. The values inside the parentheses represent the uncertainties in the last  two digits of the best-fitted values within 95\% confidence intervals.}
	\begin{tabular*}{1\textwidth}{@{\extracolsep{\fill}} c| c c c| c c c}
		\hline\hline
		States& $(0,0,1)$ & $\frac{1}{\sqrt{2}}(1,0,1)$ & $\frac{1}{\sqrt{3}}(1,1,1)$ & $s=0.885$ & $s=0.674$ & $s=0.469$\\
		\hline
		coll-exp & $1.047(53)$ & $1.002(09)$ & $1.004(12)$ &$1.028(40)$&$1.025(25)$& $1.019(37)$ \\
		coll-ideal& $0.999(20)$  & $0.977(20)$ & $1.004(16)$ &$1.036(28)$&$1.031(14)$&$1.016(15)$\\
		MUB&        $1.008(11)$ & $0.583(29)$ & $0.571(29)$&$0.849(86)$&$1.045(36)$&$1.074(27)$\\
		adaptive& $0.970(28)$  & $0.850(54)$ & $0.873(57)$ &$0.986(90)$&$1.058(25)$&$1.066(33)$\\
		\hline\hline
	\end{tabular*}
\end{table}

\section{Implementation of the maximum likelihood estimation}
Consider quantum state tomography with a POVM $\{E_j\}_{j=1}^K$ composed of $K$ outcomes. If the quantum system is characterized by the state $\varrho$, then the probability of obtaining outcome $j$ is $p_j=\tr(\varrho E_j)$. Suppose the POVM is performed $N$ times on $N$ identically prepared quantum systems, and outcome $j$ occurs $n_j$ times with $\sum_j n_j=N$. Now our task is to infer the state of the quantum system from the measurement data $D=\{n_1,n_2,\dots,n_K\}$.

As a popular estimation strategy in quantum tomography, the ML estimation~\cite{PariR04Book} searches for the quantum state $\hat{\varrho}_{\mathrm{ML}}$ that maximizes the likelihood function, i.e.,
\begin{equation}\label{eq:likelihood}
\hat{\varrho}_{\mathrm{ML}}:=\arg\max_{\varrho}L(D|\varrho),\quad\mbox{with}\;
L(D|\varrho)=\prod_j p_j^{n_j}.
\end{equation}
In practice, it is more convenient to work with the normalized log-likelihood function defined as $\cF(\varrho):=\frac{1}{N}\ln L(D|\varrho)$.
In quantum state tomography with individual measurements, the function  $\cF(\varrho)$ is concave in $\varrho$ and thus has a unique maximum  in  the quantum state space, which is convex. In addition, iterative algorithms can  be employed by following the gradient
\begin{equation}
G(\varrho)=\sum_j \frac{f_j}{p_j}E_j
\end{equation}
of $\cF(\varrho)$, where $f_j=n_j/N$ is the relative frequency.  In quantum state tomography with two-copy collective measurements as considered in this work,
however, two-copy quantum states comprise only a subset of the two-qubit state space. Therefore, standard  ML algorithms do not apply directly.

Recently, a new optimization strategy, i.e., the accelerated projected-gradient  (APG) method was introduced  in quantum tomography \cite{apg}, using which all constraints can be cast into a projection operation. In the current scenario, we have to make sure that the update for $\varrho$ at each iterative step takes on the form $\varrho=\rho^{\otimes2}$. To this end, we introduce the projection operation $\mathcal{P}$  as follows,
\begin{equation}\label{eq:proj}
\tilde{\varrho}={\cal P}(\varrho): \,\,\tilde{\varrho}=\tilde{\rho}^{\otimes2} \,\,\mbox{with}\,\, \tilde{\rho}:=\arg\min_{\rho}{||\varrho-\rho^{\otimes2}||_{\mathrm{HS}}}\,,
\end{equation}
where $||\cdot||_{\mathrm{HS}}$ denotes the Hilbert-Schmidt norm. This optimization can be  done easily by properly parametrizing  single-qubit states. Then, we modify the APG algorithm presented in \rcite{apg} as follows:
\begin{algorithm}[H]
\caption{{\bf APG for collective measurements}}
\begin{algorithmic}
\vspace*{0.1cm}
\State Given $\epsilon>0$ and $0<\beta<1$.
\State Initialize with any state $\varrho_0=\rho_0^{\otimes2}$, $\cF_{0}=\cF(\varrho_0)$;
set $\tau_0=\varrho_0$, $\theta_0=1$.
\vspace*{0.1cm}
\For {$k = 1,2,\cdots,$}
\vspace*{0.1cm}
\State Update $\varrho_k={\cal P}[\tau_{k-1}+\epsilon\,G(\tau_{k-1})]$ according to Eq.~\eqref{eq:proj}, $\cF_{k}=\cF(\varrho_k)$.
\vspace*{0.1cm}
\State Termination criterion!
\vspace*{0.1cm}
\If {$\cF_{k}<\cF_{k-1}$} \hspace*{0.1cm} (Restart)
\State Reset $\epsilon=\beta\epsilon$, $\varrho_k=\varrho_{k-1}$, $\tau_k=\varrho_k$, and $\theta_k=1$.
\Else \hspace*{0.1cm}(Accelerate)
\State Set $\theta_k=\tfrac{1}{2}{\left(1+\sqrt{1+4\theta_{k-1}^2}\,\right)}$,
then update $\tau_k=\varrho_k+\frac{\theta_{k-1}-1}{\theta_k}\left(\varrho_k-\varrho_{k-1}\right)$.
\EndIf
\vspace*{0.1cm}
\EndFor
\end{algorithmic}
\end{algorithm}

\begin{figure*}
	\center{\includegraphics[scale=0.5]{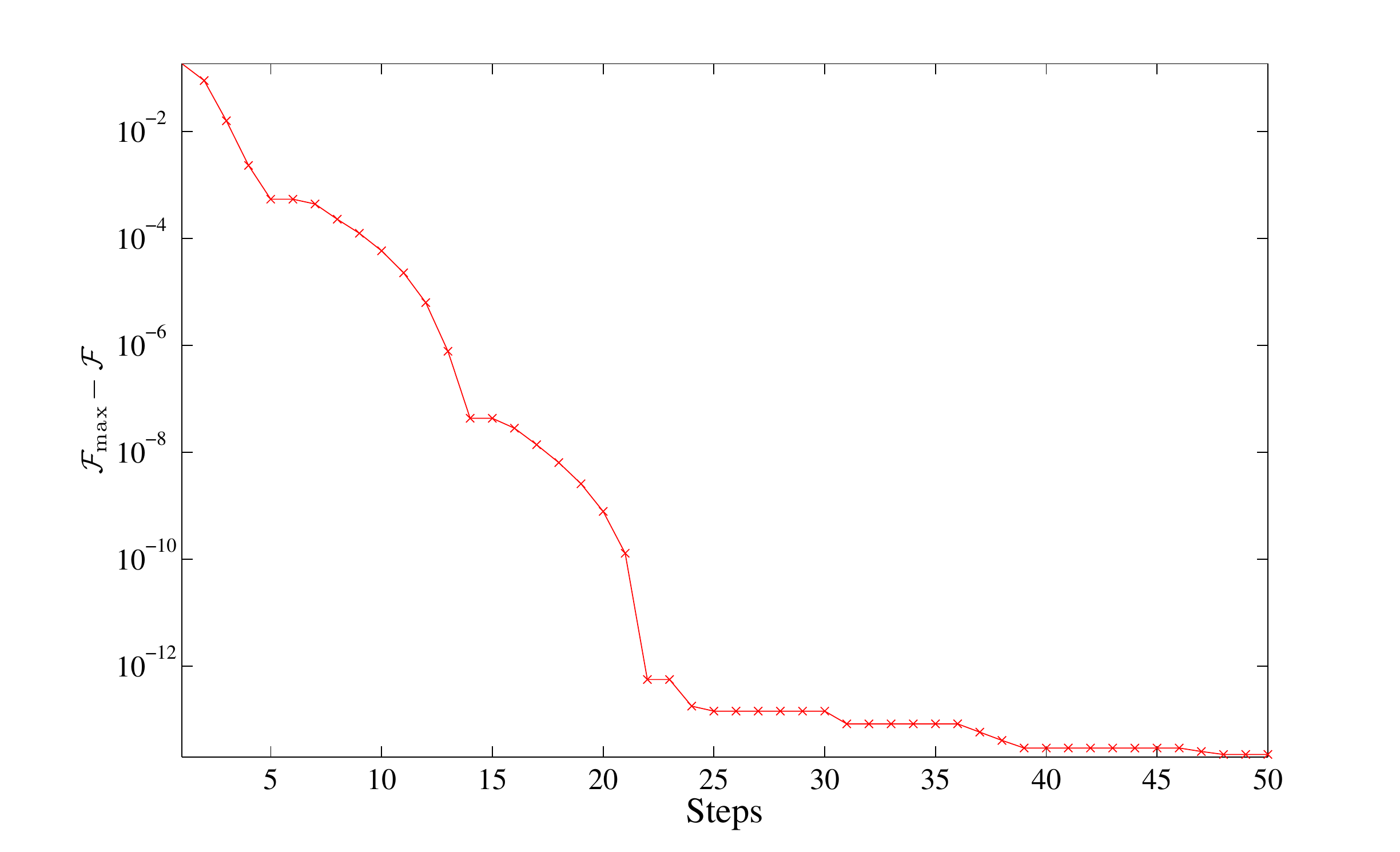}}
	\caption{\label{fig:convergence}
		Convergence of the APG algorithm in quantum sate tomography with the  collective SIC-POVM. 
		The vertical axis represents the deviation $\cF_\mathrm{max}-\cF$ of the normalized log-likelihood $\cF$ at each iterative step from the maximum value $\cF_{\max}$. In the numerical simulation,  the frequencies of obtaining the five outcomes are set to  the corresponding probabilities when the  collective SIC-POVM is performed on a two-copy state.
		The APG algorithm is run until the machine precision is reached. The figure shows that the algorithm converges very quickly.
	}
\end{figure*}
Generally speaking, the APG algorithm works in a similar way to those conventional gradient approaches, but with a tweaked gradient direction in each step to boost the convergence. Specifically, each update of the target $\varrho$ in APG is based on another state $\tau$, which gives each update some ``momentum'' from the previous step. The momentum is controlled by the parameter $\theta$, which is reset to 1 whenever it causes the current step to point too far from the direction specified by $G(\cdot)$. Upon convergence, $\varrho$ and $\tau$ will eventually merge to the same point. For more technical details about the APG algorithm, e.g., the `Restart' and `Accelerate' operations, see \rcite{apg} and references therein.

\Fref{fig:convergence} illustrates the convergence of 
the APG algorithm applied to the collective SIC-POVM.  In the numerical simulation, a qubit state $\rho$ is   generated randomly, and the frequency of obtaining each outcome is set to  the corresponding probability, that is,  $f_j=p_j=\tr(\rho^{\otimes2 }E_j)$, where $E_j$ for $j=1$ to 5 are the five outcomes of the collective SIC-POVM. In this example, the maximum value $\cF_{\max}$ of the normalized log-likelihood function is attained at the true state $\rho$; the deviation $\cF_{\max}-\cF$ from the maximum value is plotted as a function of the number of steps. 
The figure shows that the APG algorithm converges very quickly.

\end{document}